\documentclass[11pt,a4paper,oneside]{article}
\usepackage[top=1in,bottom=1in,left=.5in,right=.5in]{geometry}
\usepackage[margin=.5in,small]{caption}

\usepackage{authblk}
\usepackage{cite}
\usepackage{amsmath,gensymb}
\usepackage{epsfig}
\usepackage{graphics}
\usepackage{amsfonts, amssymb, upgreek}
\usepackage{textcomp}
\usepackage[english]{babel}
\usepackage{blindtext}
\usepackage{color}

\providecommand{\abs}[1]{\vert #1\vert}

\newcommand{\const}{\mathop{}\!\mathrm{const}}
\newcommand{\loc}{{\mathop{}\!\mathrm{loc}}}

\newcommand{\ex}{\mathrm{e}}
\newcommand{\ic}{\mathrm{i}\,}

\begin{document}


\title{Intrinsic anomalous scaling in a ferromagnetic thin film model}

\author[1]{M. F. Torres}
\author[1,2]{R. C. Buceta\thanks{rbuceta@mdp.edu.ar}}

\affil[1]{Departamento de F\'{\i}sica, FCEyN, Universidad Nacional de Mar del Plata}
\affil[2]{Instituto de Investigaciones F\'{\i}sicas de Mar del Plata, UNMdP and CONICET}
\affil[{ }]{Funes 3350, B7602AYL Mar del Plata, Argentina}

\maketitle

\abstract{
Recently, the interest on theoretical and experimental studies of dynamic properties of the magnetic domain wall (MDW) of ferromagnetic thin films with disorder placed in an external magnetic field has increased. In order to study global and local measurable observables, we consider the $(1+1)$-dimensional model introduced by Buceta and Muraca [Physica A 390 (2011) 4192], based on rules of evolution that describe the MDW avalanches. From the values of the roughness exponents, global $\zeta$, local $\zeta_\text{loc}$, and spectral $\zeta_s$, obtained from the global interface width, hight-difference correlation function and structure function, respectively, recent works have concluded that the universality classes should be analyzed in the context of the anomalous scaling theory. We show that the model is included in the group of systems with intrinsic anomalous scaling ($\zeta\simeq 1.5 $, $\zeta_\text{loc}=\zeta_s\simeq 0.5 $), and that the surface of the MDW is multi-affine. With these results, we hope to establish in short term the scaling relations that verify the critical exponents of the model, including the dynamic exponent $z$, the exponents of the distributions of avalanche-size $\tau$ and -duration $\alpha$, among others. 
}


 
\section{Introduction}
\label{sec:intro}

Due to internal disorder, the dynamic properties of the domains of ferroic materials (ferro-magnetic, -electric, or -elastic) is dominated by inhomogeneities. This disorder includes vacancies, defects, impurities and dislocations, among others. When an external driving field is present, the disorder fixed on the material limits the movement of the interfaces or domain walls separating domains. This disorder, quenched in the ferromagnetic media, is called Barkhausen noise \cite{Barkhausen-19}. Each magnetic domain wall (MDW) exhibits jerky movements between multi-metastable states of quiescence, known as Barkhausen  av\-al\-anches or jumps. Thermal fluctuations favor the elastic displacements of the MDW and eventually take it out of a quiescent state, but it can only be moved by the external field. Thus, locally the MDW can leave a metastable state by two competitive mechanisms: thermal fluctuations or driving fields. 

The study of MDW avalanches has become relevant in the field of memory devices \cite{Parkin-08,Allwood-05,Beach-08}, nanowires \cite{Singh-10,Im-09,Yamaguchi-04,Hrkac-11}, and metal, alloy or semiconductor thin-films \cite {Balk-11,Metaxas-07,Yamanouchi-07,Lemerle-98}. When ferromagnetic (or ferroelectric) thin-films with quenched disorder are placed in an external magnetic (or electric) field, a rich phenomenology is observed through the study of dynamic properties in the criticality \cite {Kim-09,Shin-07,Dourlat-08}. An important feature of the aval\-anches in materials with self-organized criticality is their scale invariance in distributions of aval\-anche-size and -duration, with power-law behaviors and well defined critical exponents. These exponents, obtained from experiments and models, fall into two classes of universality. 
One class includes materials where the dipolar interactions (or long-range) are dominant. 
The other class, in contrast, includes materials where exchange interactions (or short-range) are dominant. 
These properties have been studied intensively in bulk materials assuming the usual scaling relationships  \cite{Durin-05,Sethna-05,Colaiori-08}. With the strong interest in the study of ferromagnetic thin films, some concepts of 3-dimensional materials started to be reviewed \cite{Sethna-07}. In thin-films, as well as in bulk materials, the movement of the MDW is also dominated by depinning, although the properties start to change as a function of the film thickness \cite{Santi-06}. Experimental results show that below the 200 nm of thickness, the exponents of avalanche-size and -duration have values different from 3-dimensional magnets, but also are included in two universality classes based on the dominant type of interactions present in the medium \cite{Magni-09,Shin-08,Ryu-07,Kim-03a,Kim-03b,Puppin-00}. The theoretical models that adequately reproduce the few available experimental results on thin-films, are based on discrete 2-dimensional models \cite{Buceta-11,Rosso-03}. However, new issues have arisen when establishing relations between different scaling exponents that characterize these systems.

The pinning-depinning transition of these systems is of second order, being the average velocity $v$ of the MDW the order parameter. The MDW at non-zero temperature can move taking multi-stable states of pinning-depinning, {\sl i.e.} states of quiescence alternated with avalanches, before reaching the phase state of moving or pinning. Until today, theoretical approaches to explain the depinning transition of the MDW into disordered medium pushed by an external magnetic field, are based on: (a) the continuous equation of Edwards-Wilkinson with quenched noise (QEW) \cite{Duemmer-05,Kolton-06a,Kolton-06b,Bustingorry-08} and (b) discrete models based on microscopic structures and interactions, such as random-field Ising field model with driving (DRFIM) \cite{Qin-12,Zhou-10,Zhou-09,Roters-01,Roters-00}.

Thin-films and 2-dimensional models can display an\-omalous behavior in some measured local observables, {\sl e.g.} local interface width or roughness, height-difference correlation functions, and structure function or power spectrum, among others. These scaling anomalies in the interface local properties cannot be concluded directly only from studies of the distributions of avalanche-size and \mbox{-duration}. Recently, a theoretical study connected the an\-omalous roughness exponents to exponents of avalanche, establishing scaling relations of general validity, and applying the theory to forced-flow imbibition fronts \cite{Lopez-10}. Systems with anomalous scaling have at least one local observable with nonstationary power-law behaviors for all time \cite{Lopez-97b}. At distances or wavelengths much smaller or much larger than the correlation length, this behavior is not necessarily the same. In contrast, systems with usual or Family-Visek scaling \cite{Family-85} show, in all local observables, nonstationary (stationary) power-law behavior for distances or wavelengths much smaller (larger) than the correlation length. Each local observable is characterized by at least two exponents. If we study the $2^\text{nd}$-order height-difference correlation or -equivalently- the local interface width, different (equal) exponents of global roughness $\zeta$ and local roughness $\zeta_\text{loc}$ lead to anomalous (usual) scaling. Additionally, if we study the structure function, also known as power spectrum, different (equal) exponents of global roughness $\zeta$ and spectral roughness $\zeta_\text{s}$, lead to anomalous (usual) scaling.
The roughness exponents $\zeta\,$, $\zeta_\text{loc}$ and $\zeta_s$ allow to classify the different models and equations that have pinning-depinning transitions according to the scaling type. It has been shown that the anomaluous roughening can take different forms~\cite{Lopez-97c,Lopez-99,Ramasco-00}. Ramasco {\sl et al.}~\cite{Ramasco-00} grouped the systems into four sets, namely:
\begin{equation*}
\begin{array}{ll}
\text{usual or Family-Vicsek scaling:}&\zeta=\zeta_s=\zeta_\text{loc}<1\\
\text{intrinsic anomalous scaling:}&\zeta\neq\zeta_s=\zeta_\text{loc}<1\\
\text{super-rough anomalous scaling:}\quad &\zeta=\zeta_s >\zeta_\text{loc}=1\\
\text{faceted anomalous scaling:}&\zeta\neq\zeta_s >\zeta_\text{loc}=1\;.
\end{array}
\end{equation*}
More recent works \cite{Qin-12,Zhou-10} have confirmed that the DRFIM belongs to a universality class with spatial multiscaling and an anomalous scaling that cannot be included in any of the previous categories\footnote{First, some authors assumed that the DRFIM belonged to the QEW universality class \cite{Rosso-03,Amaral-95} and later, other authors, claimed that it had intrinsic anomalous scaling \cite{Zhou-09}.}~\cite{Chen-10}, since $\zeta\neq\zeta_\text{loc}\neq\zeta_s$ and $\zeta_\text{loc}\neq 1$. On the other hand, the QEW universality class is included in the group that has spatial single-scaling and super-rough anomalous scaling \cite{Pradas-08,Pang-00,Rosso-01,Jost-97,Jost-98,Rosso-03}.
An interface that evolves according to QEW equation is a single-valued elastic string \cite{Kolton-05,Braun-05,Kleemann-07}. 
In contrast, the interface of DRFIM is not single-valued as a result of the islands and overhangs left behind by the advancing interface \cite{Urbach-95,Amaral-95,Roters-00}. Until today, we have not identified experimental contributions that report anomalous local properties in the domain wall of ferromagnetic thin-films with Barkhausen effect. However, recent experimental contributions have shown anomalous scaling in the surface growth of films ({\sl e.g.} semiconductor and oxides) \cite{Nascimiento-11,Mohanty-11,Nabiyouni-09,Mata-08,Fu-08}.

We consider here the evolution model based on rules introduced by Buceta and Muraca \cite{Buceta-11}, and we show that it is included in an universality class with intrinsic anomalous scaling. This model takes into account the structure and exchange interactions which are present in ferromagnetic thin-films with disorder, whose MDW is driven by the external field. Its results for the distribution of avalanche-size and -duration are in agreement with experimental data. From the study of the global interface width $W$ we conclude that Family-Vicsek scaling is verified at the  criticality. We confirm that the pinning-depinning transition is clear on the saturation of the $W$. By studying the height-difference correlation function up to $4^\text{th}$-order, we show that the MDW monolayer is multi-affine.
As the most important result of this work, we corroborate that the model is capable of predicting dynamic anomalies in the local properties of the MDW, that have not been experimentally observed but have been predicted theoretically by other models. We show first that the local roughness exponent of the height-difference correlation function of second order is different to the global roughness exponent obtained from the global interface width. This indicates that the Family-Vicsek scaling is not appropriate and that an anomalous behavior is present. To complete the analysis, we show that the anomalous scaling is intrinsic by determining the spectral roughness exponent, which coincides with the local roughness exponent. Finally, we conclude with a discussion about the presented results, our contribution to the current state of the subject and the prospects opened by this work.

\section{The model}
\label{sec:model}

We consider a piece of the ferromagnetic thin-film which includes two magnetic domains separated by a MDW. On each side of the wall, we assume opposite macroscopic magnetizations in the easy direction. We suppose that the medium is composed of magnetic dipoles and randomly distributed point defects, both arranged in the nodes of a 2-dimensional square lattice of edge $L$. In this model we assume that the defects are isolated, {\sl i.e.} two defects cannot be first-neighbors to each other, and the wall in its movement does not include defects. The structure of the MDW is considered to be merely a monolayer of dipoles with perpendicular direction to the easy direction. The model only takes into account exchange interactions between nearest-neighbor dipoles. To simulate the movement of the MDW, the lattice is assumed to have periodic boundary conditions. The point defects are represented by a random pinning force $\eta(i, j)$, uniformly distributed in $[0, 1]$, assigned to each node $(i, j)$ of the lattice. Taking a lattice with density $p$ of dipoles and \mbox{$1-p$} of defects, if $\eta (i, j) < p$ the node $(i, j)$ has a dipole, otherwise it has a point defect. We characterize the disorder of the lattice using the function $F (i, j) = \Theta (p - \eta(i, j))$, where $\Theta(x) = 1$ if $x\ge 0$ and $\Theta(x) = 0$ if $x < 0$. Since the MDW is a monolayer of dipoles, a dipole located at a node $(k, n_k)$ inside the MDW ($k = 1,\dots , L$) is described by $F(k,n_k)=1$. A dipole (or defect) outside the MDW is described by $F (k,\ell) = 1\,(\text{or}\,0)$ with $\ell\neq n_k$.

The evolution rules include only ferromagnetic exchange (or short range) interactions, taking into account the balance of the magnetic moment of each side of the MDW in a neighborhood of the point it moves. When there is a local unbalance in the opposite direction to the movement, the MDW searches the equilibrium with probability $c$. However, if there is balance, with probability $1 - c$, the MDW can only move if there is an external force which can remove it from its metastable state. We start the Monte Carlo simulation with a flat wall, {\sl i.e.} initial condition $n_i = 1$ for all $i$. We introduce the relative position of neighbor nodes with respect to the node of the selected column: $x_j = n_{j+1} - n_j\,$ and $\,y_j = n_{j-1} - n_j\,$. A randomly chosen node $(j, n_j)$ of the MDW evolves in the lapse $\delta t$ according to the following rules:
\begin{enumerate}
\item[I.] With probability $c$, if $F (j, n_j +1) = 1$ and (a) $x_j + y_j \ge 2$ the selected node is moved one unit, {\sl i.e.} $\delta n_j = 1\,$, or (b) otherwise the selected node is pinned.
\item[II.] With probability $1-c$, if $F (j, n_j +1) = 1$ and 
(a) $x_j = y_j \ge 0$ the selected node is moves one unit above its neighbors, {\sl i.e.} $\delta n_j = x_j +1$, or 
(b) otherwise the selected node moves (or not) to the maximum between the neighbors and the same, {\sl i.e}. $\delta n_j=\max(x_j , y_j , 0)$. 
Also with probability $1 - c$, if $F (j, n_j + 1) = 0$ and 
(a) $x_j = y_j \ge 2$ the selected node is moved one unit above its neighbors, {\sl i.e.} $\delta n_j = x_j + 1$, or 
(b) $x_j \neq y_j$ and $\max(x_j , y_j ) \ge 3$ the chosen node moves to the maximum among its neighbors, {\sl i.e.} $\delta n_j = \max(x_j , y_j )$, or (c) otherwise the selected node does not move.
\end{enumerate}
As the MDW consist of dipoles, a rule is frustrated when a wall point tries to reach nodes with defects.

\section{Scaling analysis and Results}
\label{sec:scaling}

\subsection{Global interface width}
\label{sec:W}

\begin{figure}
\vspace*{.5cm}
\centerline{\includegraphics[width=.66\columnwidth]{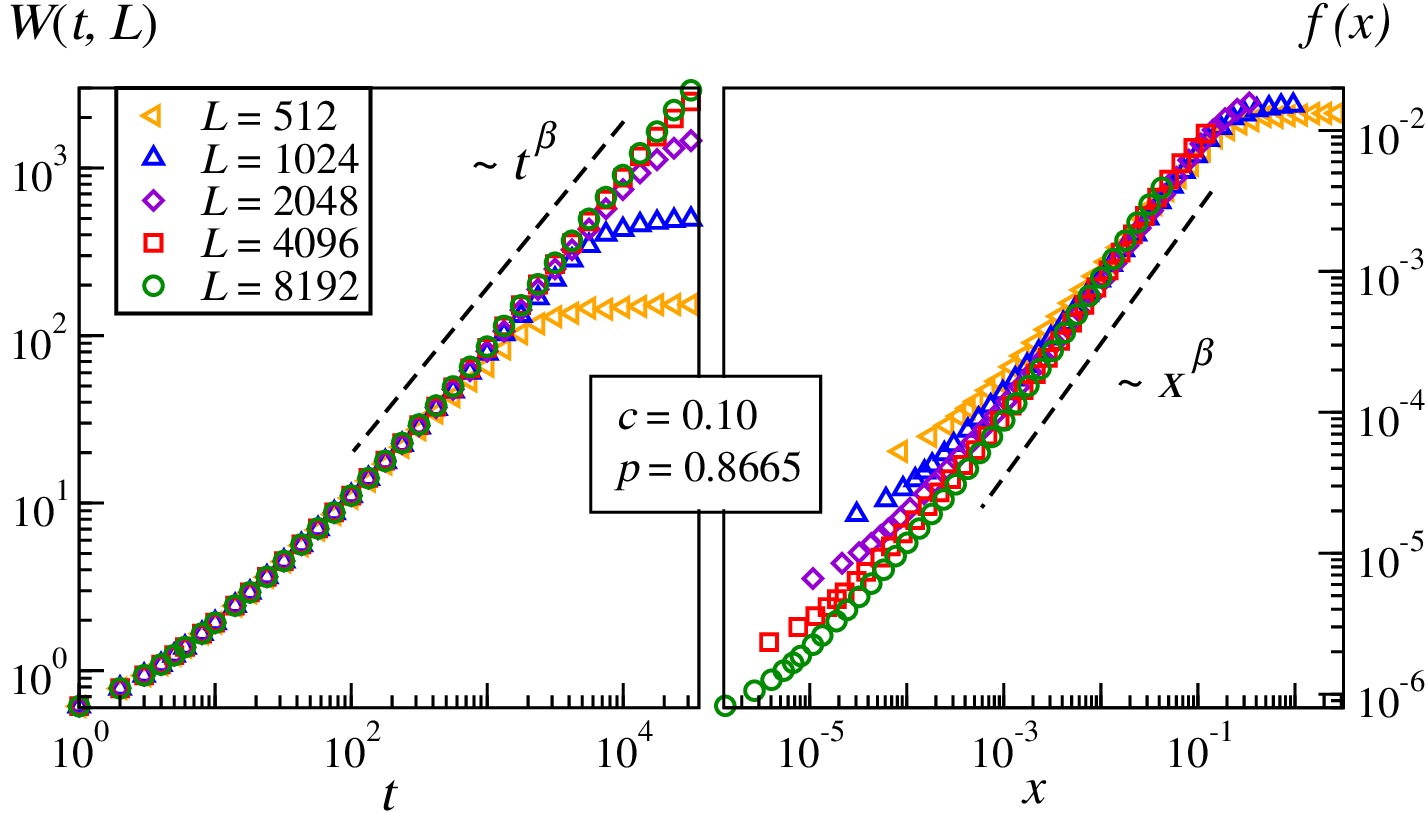}}
\caption{(Color online) Left: Plot of the global interface width $W (t, L)$ as a function of time $t$ at several lattice size $L$. The simulation results for $c = 0.10$ and $p = 0.8665$ close to critical value are shown. The dashed line shows the dynamical behavior $W\sim t^\beta$ for $t\ll L^z$. Right: Plot of the scaling function $f(x)$ {\sl vs} $x$, obtained {\sl via} Family-Vicsek scaling [eq.~(\ref{eq:Wscal})] with the left-graph data, taking $z = \zeta = 1.5\;$. We see that the points that do not fit the scaling correspond to the early regime.}\label{fig1}
\end{figure}
\begin{figure}
\vspace*{1cm}
\centerline{\includegraphics[height=.34\columnwidth]{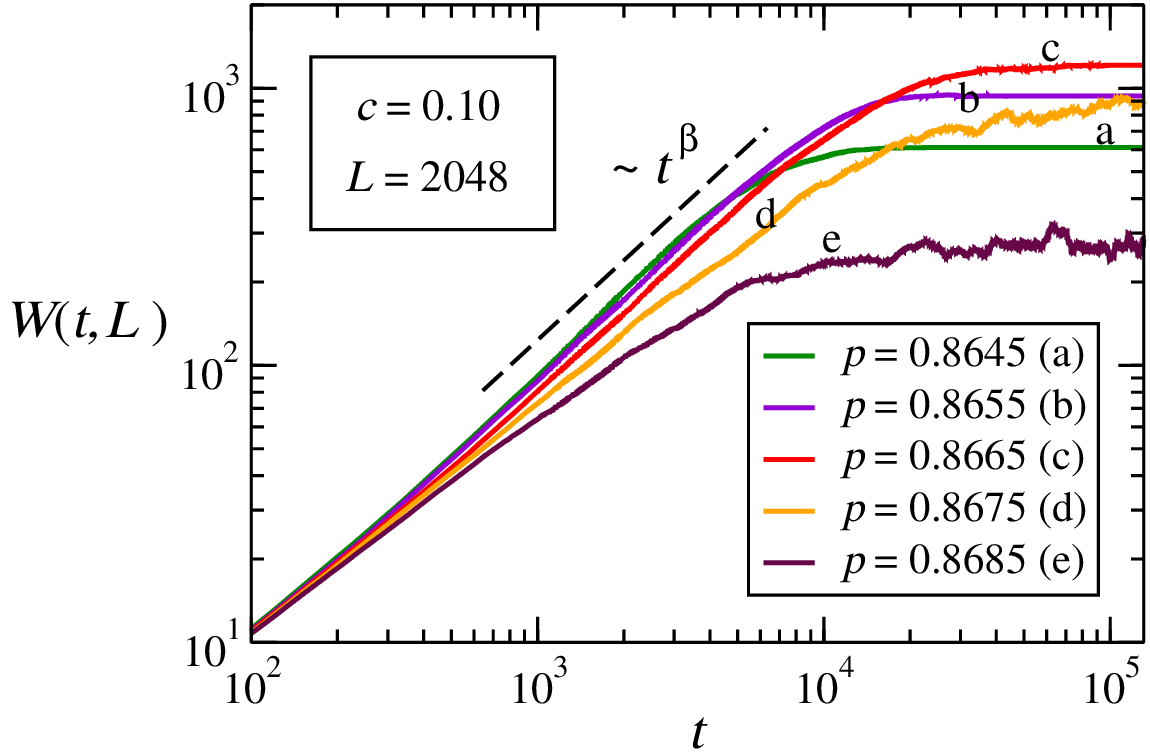}}
\caption{(Color online) Plot of the global interface width $W (t, L)$ as a function of time $t$ with lattice size $L = 2048$ for several values of $p$ around the critical value $p_c\simeq 0.8665$ and $c = 0.10$.
We notice that $W$ for $t\gg L^z$ saturates with (without) temporal fluctuations above (below) the critical value.}\label{fig2}
\end{figure}
\begin{figure}
\vspace*{1cm}
\centerline{\includegraphics[height=.34\columnwidth]{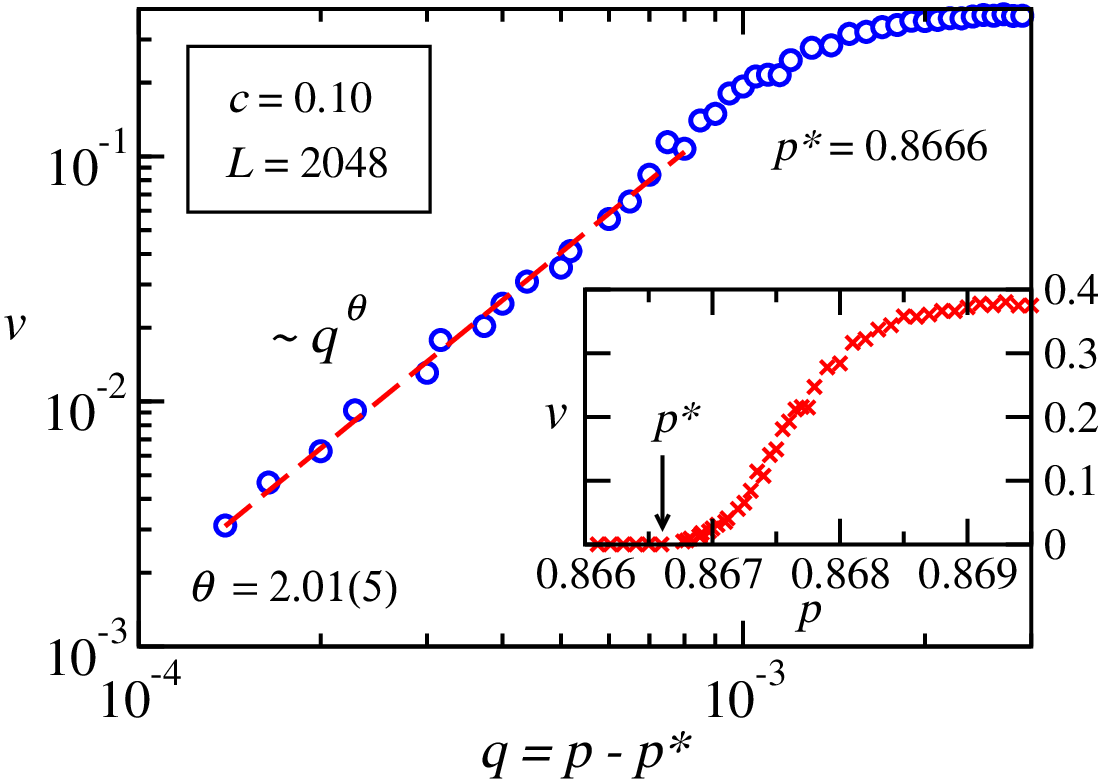}}
\caption{(Color online) Log-Log plot of the average velocity $v$ as a function $p-p^*$ with lattice size $L = 2048$ and $c = 0.10$ for $p>p^*=0.8666$ close to the critical value $p_c$, and temporal data where $W$ saturates ($t>10^5$). We see the characteristic power-law behavior of $v$ close to criticality, {\sl i.e.} $v\sim \abs{p-p^*}^\theta$ with $\theta=2.01(5)$. Inside: linear plot of $v$ as a function of $p$, with the same simulation data. We notice that $v(p)=0$ for all $p\le p^*$.}\label{fig3}
\end{figure}

In order to determine the roughness properties of the MDW, we first study the global interface width $W$ as a function of lattice size $L$ and time $t$, defined as
\begin{equation}
W(t,L)=\bigl\{\bigl\langle[n_j(t)-\langle n_j(t)\rangle_L]^2\bigr\rangle_L\bigr\}^{1/2}\;,
\end{equation}
where $\langle\cdots\rangle=\frac{1}{L}\sum_{j=1}^L\cdots$ is the spatial average over the system size $L$ and $\{\cdots\}$ is the sample ensemble average. Figure~\ref{fig1}~(left) shows $W$ as a function of $t$, for different values of $L$, close to critical value $p_c$. After the early time, the data overlaps much before the crossing time $t_\text{x}=L^z$ with dynamical behavior $W\sim t^\beta$, {\sl i.e.} for $t\ll t_\text{x}$. In contrast, $W\sim\const$ for $t\gg t_\text{x}$ and the saturation value is function of size $L$. This growth process shows the usual dynamic scaling proposed by Family-Vicsek \cite{Family-85}
\begin{equation}
W(t,L)=L^\zeta\,f(t/L^z)\;,\label{eq:Wscal}
\end{equation}
where $\zeta$ is the global roughness exponent, and $z$ is the dynamical exponent. The scaling function $f$ is 
defined by
\begin{equation}
f(x) \sim \left\{
\begin{array}{ll}
x^\beta & x\ll 1\\
\const\quad & x\gg 1\;,
\end{array}
\right.\label{eq:fscal}
\end{equation}
with the growth exponent $\beta$ linked to other exponents by the scaling relation $\zeta = z\beta$. Figure~\ref{fig1}~(right) shows the scaling function $f(x)$ and we observe the Family-Vicsek scaling for $z=\zeta=1.5\,$. The graphic shows the behavior given by eq.~(\ref{eq:fscal}). We notice that the slope of dashed line $\beta = 1$ is reached with $L\rightarrow +\infty$. 
Figure~\ref{fig2} shows the change in the saturation behavior of $W$ in the pinning-depinning transition near $p_c\simeq 0.8665$. We plot $W$ as a function of $t$, with $L$ fixed, for several values of $p$ close to $p_c$. Below criticality $W$ saturates at a constant value, where the average velocity of the wall is equal to zero. In contrast, above criticality  $W$ has temporal fluctuations around the constant value of saturation, where the average velocity of the wall is not zero. Thus for $p\lessgtr p_c$ there is a pinning or moving phase, respectively. Figure~\ref{fig3} shows the average velocity around the pinning-depinning transition for $t\gg t_\text{x}$. The velocity decreases to zero nearby above the threshold $p^*\simeq p_\text{c}$ as a power law $v\sim\abs{p-p^*}^\theta$, where $\theta$ is the velocity exponent. 
Simulations with fixed size $L$ and different values ​​of $c$, let us see that the values of the threshold ​​$p$ and exponent $\theta$ are smooth functions of the parameter $c$. This behavior, characteristic of this model, requires a particularized study that enables to compare it with other models.
Far above criticality the velocity $v\sim p$, a characteristic of quenched models which should be studied in detail beyond the present work.

\subsection{Height-difference correlation functions}
\label{sec:Gm}

\begin{figure}
\vspace*{1cm}
\centerline{\includegraphics[height=.34\columnwidth]{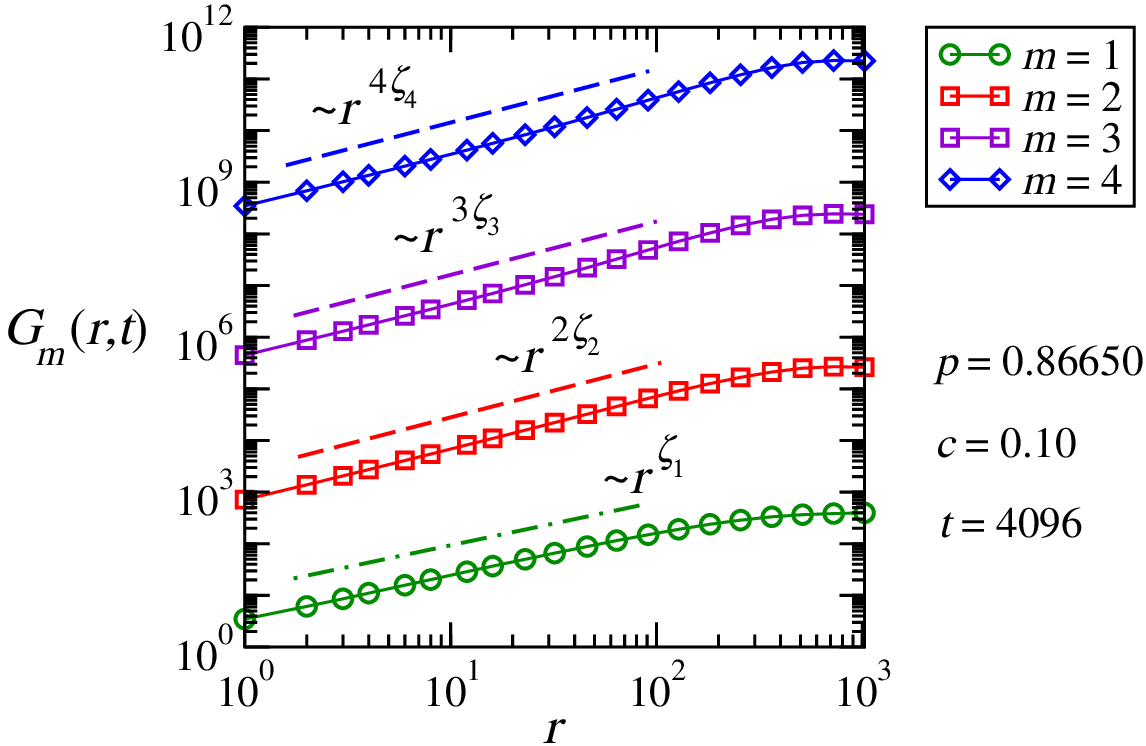}}
\caption{(Color online) Plot of the $m^\text{th}$-order height-difference correlation $G_m(r,t)$ as a function of distance $r$ at time $t = 4096$, with $m \le 4$. We see the power-law behavior $G_m \sim r^{m\zeta_m}$ for $r\ll t^{1/z}$. This plot shows results for $c = 0.10$ and $p = 0.8665$ close to critical value. The slopes are: $\zeta_1 = 0.844(4)$, $2\zeta_2 = 1.004(4)$, $3\zeta_3 = 1.036(9)$, and $4\zeta_4 = 1.045(8)$.}\label{fig4}
\end{figure}
The movement of MDW, as other surface growth processes, can be studied by the set of height-difference correlations functions of $m^\text{th}$-order
\begin{equation}
G_m(r,t)=\left\{\bigl\langle\abs{n_{j+r}(t)-n_j(t)}^m\bigr\rangle_{\!L}\right\}\;.\label{eq:Gm}
\end{equation} 
First we consider the correlation $G_m$ with fixed time $t$. We expect a power-law behavior as a function of distance $r$, {\sl i.e.} $G_m \sim r^{m\zeta_m}$ for $r\ll\xi(t)\sim t^{1/z}$, where $\xi$ is the time-dependent correlation length. If $\zeta_m$ depends on $m$ the surface is multi-affine. Otherwise, if $\zeta_m = \zeta$ the surface is self-affine. $G_m$ is expected to saturate for $r\gg \xi$. Figure~\ref{fig4} shows the correlation $G_m$ as a function of distance $r$ at fixed time $t$ for the four first values of $m$. We observe that $\zeta_m\simeq 1/m$ for $m \ge 2$ and $\zeta_1\simeq 0.84$, which ensures that the surface of the MDW is multi-affine. The decrease of these roughness exponents with the order $m$ is evidence of spatial multiscaling, which usually involves anomalous scaling \cite{Fu-08}. In addition, the relation between anomalous scaling and multiscaling has been studied in detail by Asikainen {\sl et al.} \cite{Asikainen-02a, Asikainen-02b} for single-valued height fronts in fractals. On the other hand, there are systems with anomalous scaling and spatial single-scaling \cite{Pradas-08}.
Also Figure~\ref{fig4} shows for $r\gg t^{1/z}$ saturation of the correlation function $G_m$. Particularly, $G_2$ must have the same behavior as $W^2$ for $\xi\ll r<L$ ({\sl i.e.} $t\ll t_\text{x}= L^z$), since $W (L, t) \sim t^\beta\sim \xi^\zeta$, with $\zeta=z\beta$. We propose a scaling
\begin{equation}
G_2(r,t)=\xi^{2\zeta}\,\mathcal{G}(r/\xi)\;,\label{eq:G2-a}
\end{equation}
where the scaling function $\mathcal{G}(x) \sim x^{2\zeta_2}$ if $x\ll 1$ and $\mathcal{G}(x) \sim \const$ if $x\gg 1$. Here $\zeta$ is the global roughness exponent, while $\zeta_2 :=\zeta_\text{loc}$ is known as the local roughness exponent. 
We should notice the following:\quad $G_2\sim r^{2\zeta_\text{loc}}\,\xi^{2(\zeta-\zeta_\text{loc})}$ for $r\ll\xi$. We have the usual scaling if $\zeta = \zeta_\text{loc}$ and, in contrast, we have the anomalous scaling if $\zeta_\text{loc} < \zeta$. For $r\gg \xi$ there is no difference between the scalings, {\sl i.e.} $G_2 \sim \xi^{2\zeta} = t^{2\zeta/z} = t^{2\beta}$. In summary, taking $\xi\sim t^{1/z}$ , we obtain
\begin{eqnarray}
G_2(r,t)\sim\left\{
\begin{array}{ll}
t^{2(\zeta-\zeta_\text{loc})/z}\,r^{2\zeta_\text{loc}}\quad &r\ll\xi\\
t^{2\zeta/z}& r\gg\xi\;.
\end{array}\right.\label{eq:G2-b}
\end{eqnarray}
\begin{figure}
\vspace*{.5cm}
\centerline{\includegraphics[width=.66\columnwidth]{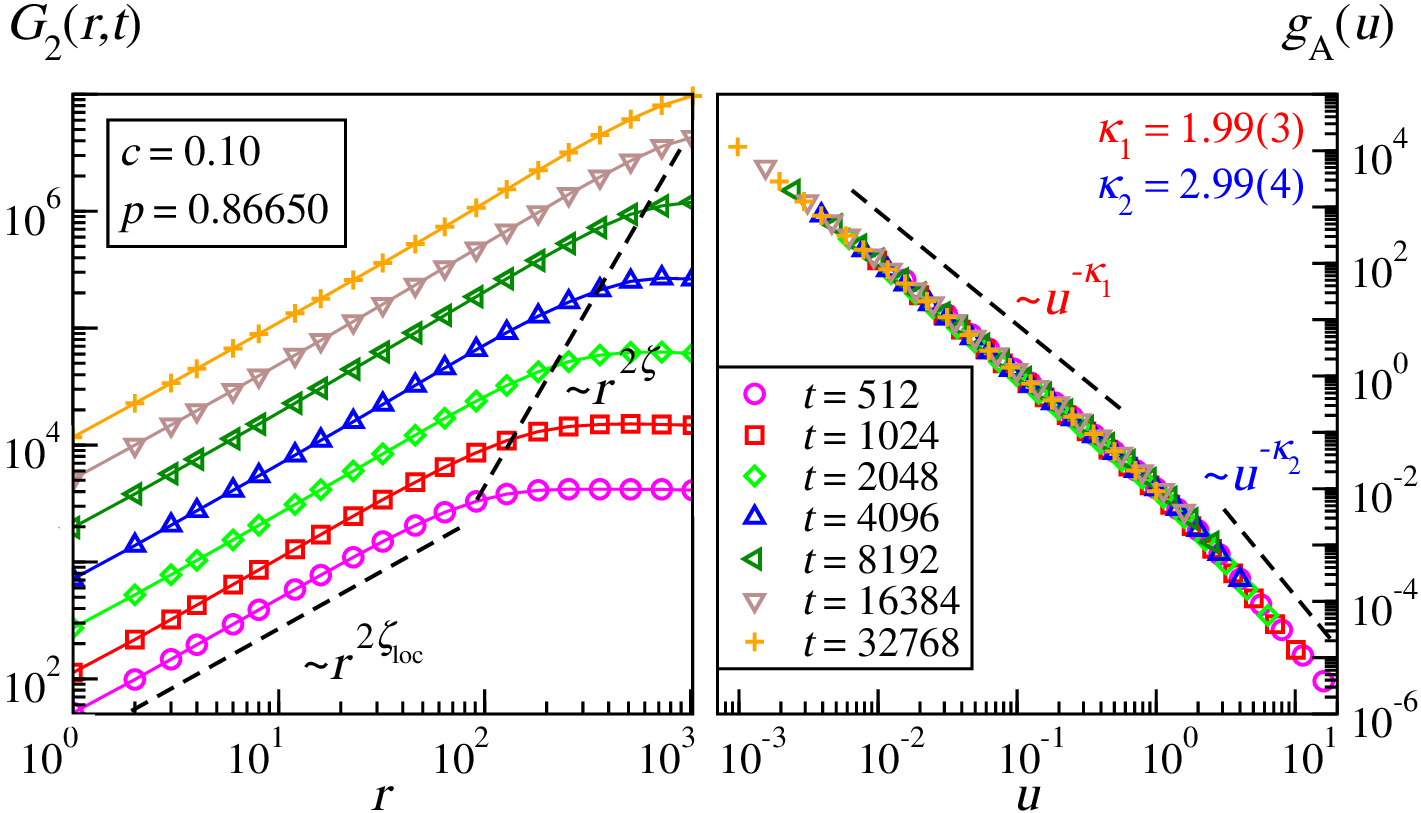}}
\caption{(Color online)  Left: $2^\text{nd}$-order height-difference correlation function $G_2(r,t)$ {\sl vs.} distance $r$  at different times $t$. The simulation results for $c = 0.10$ and $p = 0.8665$ close to critical value can be observed. We take the lattice edge $L = 1024$. For $r\ll\xi$, we observe that $G_2\sim r^{2\zeta_\text{loc}}$ and that the curves are parallel between them, which is characteristic of anomalous behavior. From eq.~(\ref{eq:G2-c}) for $r = \xi$ we conclude that $G_2\sim r^{2\zeta}$ as this plot shows. Right: Plot of the anomalous scaling function $\mathrm{g}_\text{A} (u) = r^{−2\zeta}\,G_2$ as a function of $u = t^{−1/z} r$, obtained from scaling of the data used in left-graph at various times. The exponents use for scaling are $\beta = 1$ and $\zeta = 1.5$. The obtained slopes ($-\kappa_1$ and $-\kappa_2$) are in agreement with this election. With these values we obtain $\zeta_\text{loc} = 0.495(20)$.}\label{fig5}
\end{figure}
\begin{figure}
\vspace{1cm}
\centerline{\includegraphics[height=.34\linewidth]{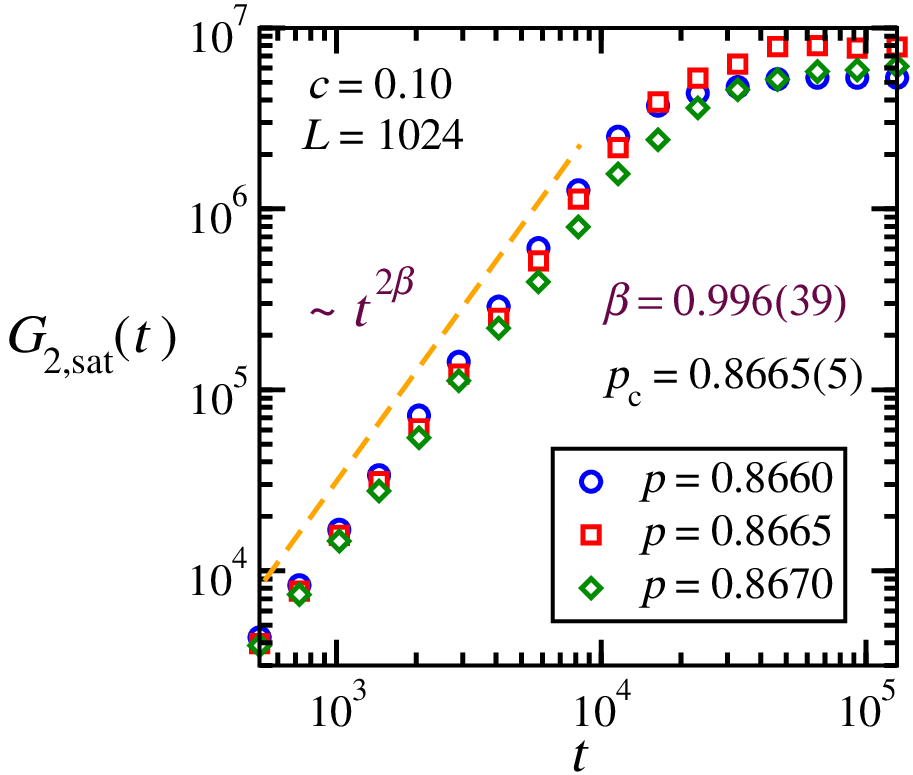}}
\caption{(Color online) The plot shows the $2^\text{nd}$-order height-difference correlation $G_2$ as a function of time $t$ for $\xi\ll r<L$ [saturation regime of Figure 3 (left)]. Results for $c = 0.10$ and values of $p$ close to estimate $p_c = 0.8665(5)$ can be observed. The data follows a power-law similar to the one which follows $W^2$ for $t\ll L^z$. The power-law fit allows to determine $\beta=0.996(39)$.}\label{fig6}
\end{figure}
\noindent 
Figure~\ref{fig5} (left) shows $G_2$ as a function of distance $r$, plotted at different times $t$. The anomalous behavior described here is observed . For $r\ll\xi$ all curves have given the same power-law behavior as $r^{2\zeta_\text{loc}}$. For $r\gg\xi$ the curves saturate at different values, as usual. We notice that the crossover points ($r = \xi$) of each curve has power-law behavior $r^{2\zeta}$. Figure~\ref{fig6} shows $G_2$ in the saturation regime as a function of time $t$. Their dynamical behavior is similar to the square interface width $W^2\sim t^{2\beta}$. The growth exponent $\beta\simeq 1$, ​​obtained from the data of Figure~\ref{fig6}, is consistent with the value ​​used from the scaling of the global interface width in Figure~\ref{fig1} (right). Equation~(\ref{eq:G2-b}) can be obtained from eq.~(\ref{eq:G2-a}) or using the scaling
\begin{equation}
G_2(r,t)=r^{2\zeta}\,\mathrm{g}_\text{A}(r/t^{1/z})\;,\label{eq:G2-c}
\end{equation}
where the so-called anomalous scaling function is
\begin{eqnarray}
\mathrm{g}_\text{A}(u)\sim\left\{
\begin{array}{ll}
u^{-\kappa_1}&\qquad u\ll 1\\
u^{-\kappa_2}&\qquad u\gg 1\;,\\
\end{array}\right.
\end{eqnarray}
being
\begin{eqnarray}
&&\kappa_1=2\,(\zeta-\zeta_\loc)\\
&&\kappa_2=2\,\zeta\;.
\end{eqnarray}
Figure~\ref{fig5} (right) shows the anomalous scaling function $\mathrm{g}_\text{A}=r^{-2\zeta}G_2$ as a function of $u=r t^{-1/z}$. Taking $z=\zeta=1.5\,$, the scaling is very good and confirms the anomalous feature of the model. We also find that $\zeta_\text{loc}\simeq 0.5$.

\subsection{Structure function}
\label{sec:S}

\begin{figure}
\vspace*{.3cm}
\centerline{\includegraphics[width=.66\columnwidth]{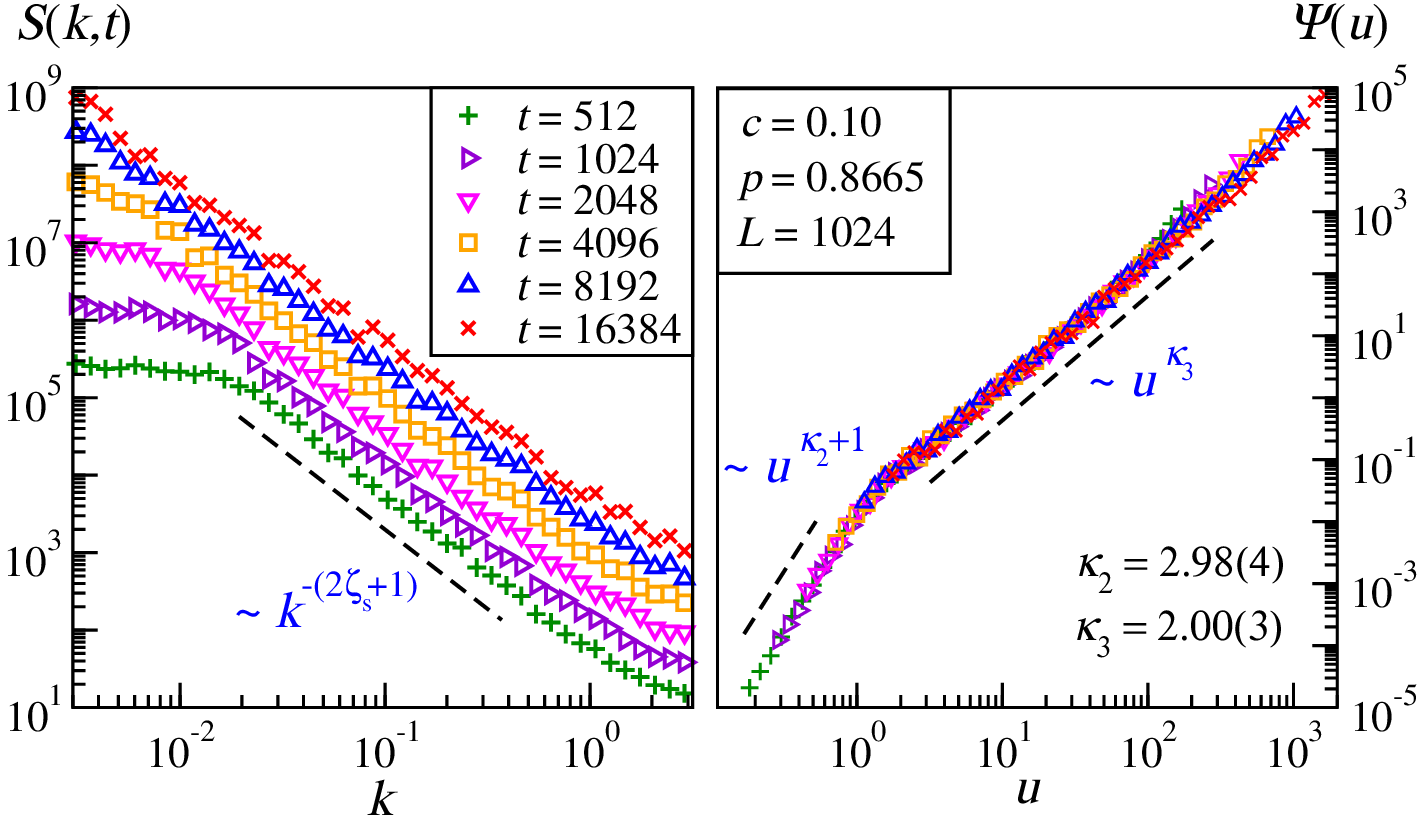}}
\caption{(Color online) Left: Plot of the spectral function $S(k, t)$ {\sl vs.} wave number $k$ at various times $t$. There is a clear power-law behavior and an anomalous character because the curves do not overlap in this regime. Right: Plot of the spectral scaling function $\Psi (u) = k^{2\zeta +1} S$ as a function of $u = t^{1/z} k$, obtained from scaling the data used in left-graph. To perform the scaling, $\zeta= 1.517$ and $\beta = 0.992$ were taken. The obtained slopes ($\kappa_2 + 1$ and $\kappa_3$) are in agreement with the selected values of $\zeta$ and $\beta$. Accordingly, $\zeta_s = 0.51(3)$ has been obtain.}\label{fig7}
\end{figure}
Instead of considering the autocorrelation function of the height, it is convenient to introduce its Fourier transform, known as the structure function or spectral power density $S(k,t)$. Introducing the Fourier transform of wall points $\tilde n_k (t) = \sum_{j=1}^N n_j(t)\,\ex^{\ic k j}$, with wave number $k = 2\pi\ic /L$ ($i = 0, 1,\dots, L-1$), the surface structure function is
\begin{equation}
S(k,t)=\left\langle \tilde n_k(t)\,\tilde n_{-k}(t)\right\rangle
\end{equation} 
Figure~\ref{fig7} (left) shows the structure function $S$ as a function of the wave number $k$ at different times $t$. We observe that for intermediate wave numbers there is a clear power-law behavior and non-overlapping curves. This is characteristic of anomalous scaling functions. It is appropriate to propose that the structure function scale as
\begin{equation}
S(k,t)=k^{-(2\zeta+d)}\,\Psi(t^{1/z}k)
\end{equation}
where the so-called spectral scaling function is
\begin{eqnarray}
\Psi(u)\sim\left\{
\begin{array}{ll}
u^{\kappa_2+d}&\qquad u\ll 1\\
u^{\kappa_3}&\qquad u\gg 1\\
\end{array}\right.\;,
\end{eqnarray}
with $d$ the surface dimension \cite{Ramasco-00} and
\begin{equation}
\kappa_3=2(\zeta-\zeta_s)\;.
\end{equation}
Figure~\ref{fig7} (right) shows the anomalous spectral scaling function $\Psi=k^{2\zeta+1}S$ as a function of
$u=k t^{1/z}$. The anomalous scaling fits very well and $\zeta_s\simeq 0.5$ has been determined, which justifies its intrinsic character.

In all figures, $c=0.10$ has been chosen arbitrarily, although other values ​​of $c\in (0,1)$ show qualitatively the same results. Furthermore, the critical value $p_c$ is a very smooth function of $c$.

\section{Conclusions}
\label{sec:conclusions}

The main contribution of this study was to characterize the statistical properties of local observables of MDW, such as the correlation and the structure function. We use a discrete rule-based model to describe the motion of the MDW between two domains of a ferromagnetic thin-film sample with defects, when placed in an external magnetic field. This 2-dimensional model has successfully described MDW avalanches \cite{Buceta-11} and has shown a pinning-depinnnig transition whose properties had not been characterized before. With the purpose of obtaining global properties we considered, as usual, the time evolution of the interface width for different lattice sizes. We observed Family-Vicsek scaling at criticality with $z=\zeta=1.5\,$. Around criticality, the global interface width in the saturation displayed the characteristic of the pinning-depinnig transition. Above (below) criticality the global interface width saturated with (without) fluctuations. We found that the interface is multi-affine by studying the height-difference correlation functions of different orders, all with power-law behaviors. We show that the $2^\text{nd}$-order correlation exponent (or local roughness exponent) is not equal to the global roughness exponent. This fact, among others outlined here, made us conclude that the scaling is anomalous. We performed the scaling with $z=\zeta = 1.5$ to determine $\zeta_\text{loc}\simeq 0.5\,$. With the purpose of classifying the anomalous scaling, we studied the structure function. In the regime of intermediate wave number, the power-law scaling was clear and allowed us to determine, with high precision, the spectral exponent and ensured that $\zeta_s=\zeta_\text{loc}$. Following the criteria introduced by Ramasco {\sl et al.} \cite{Ramasco-00}, we were able to affirm that our model belongs to the set of systems that have intrinsic anomalous scaling. 
We have carried out a rigorous analysis to ensure the existence of intrinsic anomalous scaling. The local and spectral roughness exponents, obtained from regressions, coincide within error range determined by the errors of the calculated exponents.
If we analyzed the rules of evolution, in detail, we would observe that the MDW is formed by terraces and plateaus. This is a feature of interfaces with intrinsic anomalous roughening. Systems with interfaces that leave islands and overhangs in their advance as the DRFIM are not included in this category. In the context of the study of the Barkhausen effect in ferromagnetic thin-films, we believe it is promising to establish the connection between roughness and avalanche exponents, following the known theoretical framework \cite{Lopez-10}. 
According to dynamic renormalization group results, the intrinsic anomalous scaling cannot occur in homogeneous ({\sl i.e.} non-disordered) and local growth models \cite{Lopez-05}. In agreement with this result, the quenched disorder in our system is responsible for intrinsic anomalous roughening. The situation is similar to what occurs in other systems with a depinning transition, as fluid imbibition in disordered media.
Finally, another outstanding point is the morphology of the domain wall from the set of local and global exponents, which should be subject of future studies for models and experiments.

\section*{Acknowledgements}

R.C.B. thanks to D. Hansmann and C. Rabini for your suggestions on the final manuscript.

\bibliography{arxiv-Torres-Buceta-2012c.bib}

\end{document}